\documentclass[aps,prl,twocolumn,reprint,showpacs,amsmath,amssymb,superscriptaddress,citeautoscript]{revtex4}
\usepackage{graphicx}

\begin{document}


\title{Charge-Doping driven Evolution of Magnetism and non-Fermi-Liquid
Behavior \\ in the Filled Skutterudite CePt$_4$Ge$_{12-x}$Sb$_x$}

\author{M.\ Nicklas}
 \affiliation{Max Planck Institute for Chemical Physics of Solids, N\"{o}thnitzer Str. 40, 01187
Dresden, Germany}
\author{S.\ Kirchner}
 \affiliation{Max Planck Institute for Chemical Physics of Solids, N\"{o}thnitzer Str. 40, 01187
Dresden, Germany}
 \affiliation{Max Planck Institute for Physics of Complex Systems, N\"{o}thnitzer
Str. 38, 01187 Dresden, Germany}
\author{R.\ Borth}
 \affiliation{Max Planck Institute for Chemical Physics of Solids, N\"{o}thnitzer Str. 40, 01187
Dresden, Germany}
\author{R.\ Gumeniuk}
 \affiliation{Max Planck Institute for Chemical Physics of Solids, N\"{o}thnitzer Str. 40, 01187
Dresden, Germany}
\author{W.\ Schnelle}
 \affiliation{Max Planck Institute for Chemical Physics of Solids, N\"{o}thnitzer Str. 40, 01187
Dresden, Germany}
\author{H.\ Rosner}
 \affiliation{Max Planck Institute for Chemical Physics of Solids, N\"{o}thnitzer Str. 40, 01187
Dresden, Germany}
\author{H. Borrmann}
 \affiliation{Max Planck Institute for Chemical Physics of Solids, N\"{o}thnitzer Str. 40, 01187
Dresden, Germany}
\author{A.\ Leithe-Jasper}
 \affiliation{Max Planck Institute for Chemical Physics of Solids, N\"{o}thnitzer Str. 40, 01187
Dresden, Germany}
\author{Yu.\ Grin}
 \affiliation{Max Planck Institute for Chemical Physics of Solids, N\"{o}thnitzer Str. 40, 01187
Dresden, Germany}
\author{F.\ Steglich}
 \affiliation{Max Planck Institute for Chemical Physics of Solids, N\"{o}thnitzer Str. 40, 01187
Dresden, Germany}

\date{\today}

\begin{abstract}
The filled-skutterudite compound CePt$_4$Ge$_{12}$ is situated close to the border between
intermediate-valence of Ce and heavy-fermion behavior. Substitution of Ge by Sb drives the system
into a strongly correlated and ultimately upon further increasing the Sb concentration into an
antiferromagnetically ordered state. Our experiments evidence a delicate interplay of emerging Kondo
physics and the formation of a local 4$f$ moment. An extended non-Fermi-liquid region, which can be
understood in the framework of a Kondo-disorder model, is observed. Band-structure calculations
support the conclusion that the physical properties are governed by the interplay of electron supply
via Sb substitution and the concomitant volume effects.
\end{abstract}

\pacs{71.27.+a,75.20.Hr}


\maketitle

The interplay of localized and itinerant degrees of freedom is at the heart of strongly correlated
systems and often results in the emergence of novel and unconventional electronic phases, like
unconventional superconductivity or quantum-critical behavior. Addressing this strong-coupling
problem remains a challenge, both experimentally and theoretically. The identification of new classes
of intermetallic compounds that allow to study this emergence in a well-characterized setting is
highly desirable.

The large family of filled-skutterudite compounds shows a wealth of topical physical phenomena
\cite{Uher01,SalesREHandbook,SatoMagHandbook,Maple.08}. The phases with general formula $MT_4X_{12}$
are built up by a rigid covalently bonded transition-metal ($T$) pnictogen ($X$) framework with
mostly ionically bonded filler atoms ($M$). The possible groundstates comprise various forms of
magnetic ordering {\it i.e.}\ itinerant ferromagnetism and local anti-ferromagnetism, heavy-fermion
behavior, superconductivity, half metallicity, and non-Fermi-liquid (NFL) behavior. This family of
materials also has interesting properties for thermoelectric applications.

Recently, new filled skutterudites $M$Pt$_4$Ge$_{12}$ have been reported with $M$ being Sr or Ba
\cite{Bauer07a,Gumeniuk08a}, rare-earth metals La, Ce, Pr, Nd, Sm, Eu
\cite{Gumeniuk08a,Gumeniuk11a,Gumeniuk10c} as well as Th or U \cite{Kaczorowski08,Bauer08c}. Several
$M$Pt$_4$Ge$_{12}$ compounds ($M$ = Sr, Ba, La, Pr, Th) become superconductors with $T_c$ up to
8.3\,K \cite{Gumeniuk08a}. For $M$ = Nd, Eu a well-localized nature of 4$f$ electrons is observed
\cite{Nicklas10b} while SmPt$_4$Ge$_{12}$ features a heavy-fermion state at low temperatures
\cite{Gumeniuk10c}. Theoretical and spectroscopic studies have shown that states at the Fermi level
($E_F$) can be attributed mainly to Ge 4$p$ electrons while Pt 5$d$ states are lying rather deep and
only partially form covalent bonds with Ge 4$p$ \cite{Rosner09a,Gumeniuk10b}. The electropositive
filler elements $M$ act as electron donors transferring charge to the [Pt$_4$Ge$_{12}$] polyanion
\cite{Gumeniuk10b}. In these systems, in the absence of strong correlations, simple band-structure
concepts can be used and have been successfully applied {\it e.g.}\ in the optimization of the
superconducting $T_c$ of BaPt$_4$Ge$_{12}$ by partial substitution of Pt by Au \cite{Gumeniuk08b}.

The role of electron-electron interaction in the Ce-based skutterudites has been of recent interest
\cite{SatoMagHandbook}. The question of how strong the electronic correlations are in these compounds
and how to influence them by charge carrier doping and/or changing the unit cell dimensions through
chemical substitution are as important as they are delicate. In this respect the filled Skutterudite
compound CePt$_4$Ge$_{12}$ is particularly attractive, since previous studies have placed this
skutterudite at the border between intermediate-valence and Kondo-lattice behavior
\cite{Gumeniuk11a,Toda08a}. The low-energy properties of this system are described by an Anderson
lattice model which, in the absence of charge fluctuations, reduces to the Kondo lattice
model~\cite{Hewson}. The nature of the quantum phase transition at the border of antiferromagnetism
in the Kondo lattice is much debated \cite{Loehneysen.07,Si.10a}. Only few examples are known where
NFL effects have been observed in the intermediate-valence regime due to antiferromagnetic (AF)
quantum-critical fluctuations, \textit{e.g.}, in $\beta$-YbAlB$_4$ \cite{Matsumoto11a} or
CeIn$_{3-x}$Sn$_x$ \cite{Lawrence79,Pedrazzini04}.

In this Letter, we show that by a suitable atomic substitution we can tune CePt$_4$Ge$_{12}$ from a
nearly intermediate-valence paramagnet through a NFL phase into an antiferromagnet of localized
Ce-4\textit{f} moments. This phase sequence driven by the chemical substitution is discussed in the
context of other NFL systems. Substitution of Ge by a larger isovalent element drives the system
toward more localization, thus leading to a reduction of charge fluctuations. The obvious choice for
this purpose is to replace or partially substitute Ge by Sn in CePt$_4$Ge$_{12-x}$Sn$_x$. This
failed, since only small amounts ($x < 0.3$) of Sn could be substituted \cite{Gumeniuktobe}. As
alternatives, Sb or As, which are significantly larger than Ge and provide one extra electron, were
explored \cite{remarkAs,Nicklas.11}. Contrary to Sn substitution, our studies demonstrate a large
range of stability of CePt$_4$Ge$_{12-x}$Sb$_x$ of $x$ up to 3.0 in the filled-skutterudite
structure.

Polycrystalline samples of CePt$_4$Ge$_{12-x}$Sb$_x$ were prepared by arc-melting stoichiometric
amounts of the constituent elements. For homogenization, the samples were re-melted several times
with a negligible mass loss. All samples were annealed at 820$^{\circ}$C for 10 days and
characterized by powder X-ray diffraction (XRD). The compositions of the observed phases were
confirmed by energy dispersive X-ray spectroscopy (EDXS). Heat capacity and electrical resistivity
experiments have been carried out utilizing a commercial measurement system (PPMS, Quantum Design)
and a dilution refrigerator (Oxford Instruments). The magnetic susceptibility was determined in a
SQUID magnetometer (MPMS, Quantum Design) equipped with a $^3$He-option (iQuantum). The electronic
structure of CePt$_4$Ge$_{12-x}$Sb$_x$ was studied using the full-potential local-orbital (FPLO)
minimum basis code (version 9.01-35-x86) \cite{Koepernik99} within the local density approximation
(LDA). In the scalar-relativistic calculations, the exchange-correlation potential of Perdew and Wang
was employed \cite{Perdew92}.

In CePt$_4$Ge$_{12}$, Sb substitution on the Ge site expands the unit-cell volume. A linear
dependence of the lattice parameter $a(x)$ in the CePt$_4$Ge$_{12-x}$Sb$_x$ series is observed (see
below). In addition to the expansion of the unit cell, Sb substitution changes the chemical potential
through electron doping.

\begin{figure}[tb!]
\centering
\includegraphics[angle=0,width=\linewidth,clip]{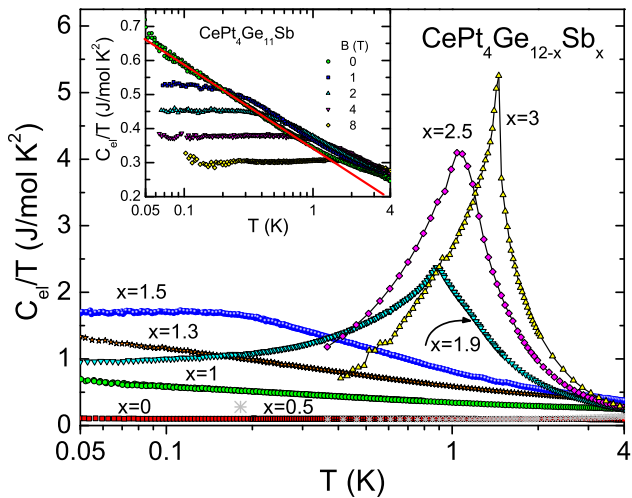}
\caption{\label{CpT} (Color online) Electronic contribution to the specific heat, $C_\text{el}(T)/T$,
of CePt$_4$Ge$_{12-x}$Sb$_x$ for different $x$. The low-$T$ nuclear Schottky contribution to $C_p(T)$
has been subtracted. Inset: $C_\text{el}(T)/T$ of CePt$_4$Ge$_{11}$Sb in different magnetic fields.
At $B = 0$, $C_\text{el}(T)/T$ exhibits a logarithmic increase toward lowest temperatures (indicated
by the red line), while $C_\text{el}(T)/T$ becomes constant in external magnetic fields.}
\end{figure}

The low-temperature properties of CePt$_4$Ge$_{12-x}$Sb$_x$ change strongly with Sb concentration.
CePt$_4$Ge$_{12}$ has been shown to be close to intermediate-valence behavior \cite{Gumeniuk11a}.
Below $T_\text{LFL} \approx 8.4$\,K CePt$_4$Ge$_{12}$ exhibits a Landau Fermi-liquid (LFL)
groundstate \cite{Gumeniuk11a}. In slightly Sb-doped samples ($x = 0.5$) the magnetic and
thermodynamic properties do not change qualitatively. The specific heat still indicates LFL behavior
for $T < T_\text{LFL} \approx 7.9$\,K. On the other side, for high Sb concentration, ($x \geq 1.9$),
we find AF order at low temperatures. In CePt$_4$Ge$_{10.1}$Sb$_{1.9}$, a sharp anomaly in the
specific heat at $T_N = 0.89$\,K indicates the magnetic ordering (see Fig.\,\ref{CpT}). Upon
increasing the Sb concentration $T_N(x)$ shifts to higher temperatures. In CePt$_4$Ge$_9$Sb$_3$, with
the highest Sb concentration of our study, $T_N = 1.46$\,K is attained. The AF nature of the ordered
phase is confirmed by magnetic susceptibility data, $M(T)/H$, and is further supported by the
magnetic-field dependence of $T_N$, which is suppressed to lower temperatures upon application of a
magnetic field (not shown).

\begin{figure}[tb!]
\centering
\includegraphics[angle=0,width=\linewidth,clip]{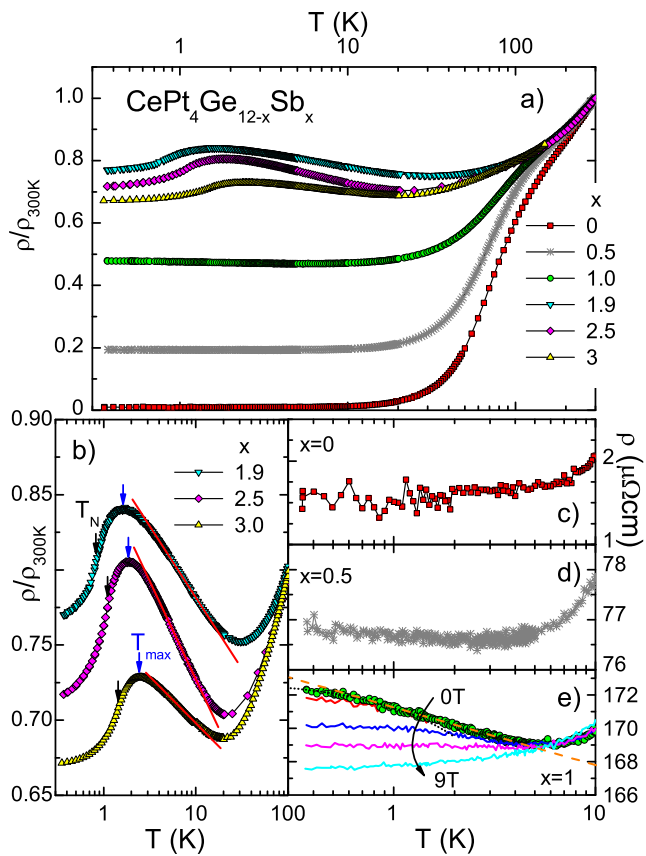}
\caption{\label{rho} (Color online) a) Normalized resistivity $\rho(T)/\rho_\text{300\,K}$ of
CePt$_4$Ge$_{12-x}$Sb$_x$ for different $x$ on a logarithmic temperature scale. b) Magnification for
$x = 1.9$, 2.5, and 3. Straight lines indicate a logarithmic dependence of
$\rho(T)/\rho_\text{300\,K}$ on $T$, reminiscent of incoherent Kondo scattering. The maximum and the
kink in $\rho(T)/\rho_\text{300\,K}$ at $T_\text{max}$ and at the N{\'e}el transition, $T_N$,
respectively, are marked by arrows. c)--e) Low-temperature resistivity $\rho(T)$ for $x = 0$, 0.5,
and 1. Except for $x = 0$ $\rho(T)$ increases toward low $T$. In e) the dashed line marks a
logarithmic and the dotted line a linear temperature dependence of $\rho(T)$. In addition to 0\,T
(circles) data in magnetic fields of 1\,T, 3\,T, 5\,T, and 9\,T (solid lines), are shown.}
\end{figure}

Magnetization measurements on the AF ordered samples ($1.9 \leq x \leq 3$) reveal that the magnetic
moment does not saturate in a magnetic field up to 7\,T. At 0.5\,K only between $0.3\mu_B/\text{Ce}$
and $0.45\mu_B/\text{Ce}$ is reached, which is below the value expected for a Ce$^{3+}$ doublet
groundstate. Even though the moment does not saturate in our experimentally accessible magnetic field
range, the obtained values suggest a saturation moment which is not compatible with the small values
typically observed in itinerant Ce systems. Also the magnetic entropy at $T_N$ is reduced compared
with $R\ln{2}$ expected for a doublet groundstate \cite{Gumeniuk11a}. From the magnetic specific
heat, we obtain $23$\%, $30$\%, and $48$\% of $R \ln 2$ at $T_N$ for $x = 1.9$, $2.5$, and $3$,
respectively. The increasing magnetic entropy recovered at $T_N$ indicates that the 4$f$ moments are
becoming more and more localized in character on increasing Sb concentration.

\begin{figure}[t!]
\centering
\includegraphics[angle=0,width=\linewidth,clip]{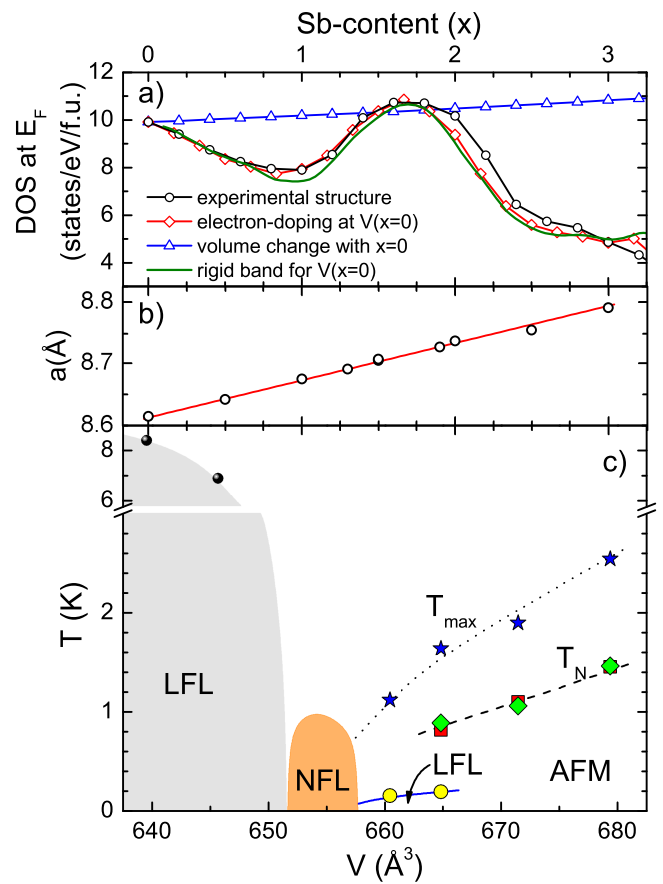}
\caption{\label{PhaseDiagram} (Color online) a) Density of states (DOS) at the Fermi energy ($E_F$)
as function of the Sb concentration $x$ calculated by different techniques/with different assumptions
as indicated. b) Experimental lattice parameter $a(x)$ of CePt$_4$Ge$_{1-x}$Sb$_x$. (c) Temperature
-- unit-cell volume ($T-V$) phase diagram of CePt$_4$Ge$_{12-x}$Sb$_x$. The corresponding Sb
concentration is indicated on the top axis. Samples $x = 0$ and 0.5 exhibit LFL behavior in an
extended temperature range. AF ordering temperature $T_N$ determined from specific heat
($\blacklozenge$) and from resistivity data ($\blacksquare$). Temperature of the maximum in the
resistivity $\rho(T)$, $T_\text{max}$ ($\bigstar$). The samples with $x = 1$ and 1.3 show NFL
behavior at low $T$. CePt$_4$Ge$_{10.5}$Sb$_{1.5}$ and CePt$_4$Ge$_{10.1}$Sb$_{1.9}$ display both LFL
behavior below $T_\text{LFL}$ ($\bullet$). The latter compound orders AF while
CePt$_4$Ge$_{10.5}$Sb$_{1.5}$ does not show any magnetic order. The regions where LFL and NFL
behavior is observed  are indicated.}
\end{figure}

To get further insight in the electronic groundstate properties of the CePt$_4$Ge$_{12-x}$Sb$_x$
series we now turn to the electrical resistivity results presented in Fig.\,\ref{rho}. The
temperature dependence of $\rho$ changes also drastically with Sb concentration. While in
CePt$_4$Ge$_{12}$ $(\rho(T)-\rho_0)\propto T^2$ indicates LFL behavior below 10\,K consistent with
the specific heat \cite{Gumeniuk11a}, in CePt$_4$Ge$_{11.5}$Sb$_{0.5}$ a tiny increase of $\rho(T)$
toward low temperatures, becomes apparent below 3\,K. This increase becomes more pronounced at $x =
1$, where $\rho(T)$ is best described by a logarithmic temperature dependence between 3.6\,K and
0.75\,K with a tendency to saturation at lowest temperatures. This contribution to the resistivity
becomes more important with increasing $x$. Furthermore, a clear maximum in $\rho(T)$ develops for $x
\geq 1.5$ (see Fig.\,\ref{rho}). Upon increasing $x$, the position of the maximum [$T_\text{max}(x)$]
shifts toward higher temperatures. In the magnetically ordered samples, $T_N$ is marked by a small
kink in $\rho(T)$ just below $T_\text{max}$. A summary of the experimental results is shown in
Fig.\,\ref{PhaseDiagram}c. The lines in this figure mark the general trend of $T_N$ and
$T_\text{max}$ in $x$ and are merely guides to the eye.

We note that both $T_\text{max}(x)$, as well as $T_{N}(x)$ seem to extrapolate to zero temperature in
the region around $x = 1$. As a result, the phase diagram of CePt$_4$Ge$_{12-x}$Sb$_{x}$ resembles at
first sight the generic phase diagram of a quantum-critical point (QCP) scenario separating a
non-magnetic LFL state \cite{Gumeniuk11a,Toda08a} on one side from an antiferromagnetically ordered
groundstate on the other side of the QCP. For $x = 1$ and 1.3 we find a logarithmic temperature
dependence of $C_\text{el}/T$ below 1\,K. The data are shown in Fig.\,\ref{CpT}. A logarithmic
divergence of $C_\text{el}/T$ with temperature is also in line with general expectations for
materials in close proximity to such AF QCPs. LFL behavior [$C_\text{el}(T)/T = \text{const.}$] is in
both samples recovered in magnetic fields and extends to higher temperatures upon increasing field,
at 8\,T to well above 1\,K. The inset of Fig.\,\ref{CpT} shows the data for $x = 1$. Yet, the phase
diagram in Fig.\,\ref{PhaseDiagram}c, reveals some features that are incompatible with the standard
AF QCP scenario: Even though the $T_N(x)$ line extrapolates to zero at $x \approx 1$, the NFL region
is separated from the magnetically ordered phase by a non-magnetic LFL. At $x = 1.5$, $\gamma =
C_\text{el}(T)/T|_{T\rightarrow0} = 1.74$ J\,mol$^{-1}$\,K$^{-2}$ becomes constant below 0.16\,K,
indicating a strongly correlated LFL groundstate, notably different from the one on the low Sb
concentration side. The $C_\text{el}(T)/T = \text{const.}$ region extends to higher temperatures upon
increasing magnetic field. Furthermore, we find no indication for a magnetic phase transition in any
probe at this Sb concentration. This extended NFL region and the separation of the magnetic phase
from the NFL behavior cannot be explained within the generic QCP scenario.

Instead, the system evolves from a metal with moderate electron correlations which is located at the
border to intermediate valence and a correspondingly high Kondo temperature into a local-moment
magnet in an unconventional fashion. A full understanding of the CePt$_4$Ge$_{12-x}$Sb$_{x}$ phase
diagram requires to disentangle the various effects of Sb substitution. Both the expansion of the
unit-cell volume as well as electron doping will reduce the strength of the Kondo interaction $J$ and
thus further localize the 4\textit{f} states as the Ce-4\textit{f} states are electron-like. Since Sb
replaces Ge, this substitution results in a strongly disordered hybridization and therefore a
distribution of the (local) Kondo temperatures: Sb substitution of the Ge atoms forming the cage
enclosing the Ce leads to significant changes in the local environment of the Ce \cite{CEFnote}. The
increase in disorder is clearly reflected in the $x$-dependence of the residual resistivity (see
Fig.\,\ref{rho}). In addition, the relevant Kondo scale, $T_K\sim e^{-1/J N_F}$, is [at least for a
structureless density of states (DOS)] also affected by the DOS at the Fermi level, $N_F$. Band
structure calculations show (see Fig.\,\ref{PhaseDiagram}a) that $N_F$ shows an overall linear
decrease with increasing $x$ with a broad maximum around $x=1.7$. A second maximum develops around
$x= 4.2$ followed by an abrupt decrease of $N_F$ down to 0 at $x\approx 5$ (not shown here), similar
to the calculations done for the system LaPt$_4$Ge$_{12-x}$Sb$_x$ in a recent study \cite{Chen.12}.
These effects result in a strong  decrease of the Kondo scale as $x$ increases (which may happen to
be somewhat slower around $x=1.5$) and a broadening in the distribution of Kondo temperatures. The
calculations based either on the experimental unit-cell volume expansion or taking the
concentration-independent $V_0$ at $x = 0$ are nearly identical and agree well with a rigid-band
approach. In this calculation, the Ce-4\textit{f} states are taken as core states. If Kondo
correlations were negligible, the Sommerfeld coefficient should mirror the behavior of $N_F(x)$ of
our band structure calculations. This is not the case (see Fig.\,\ref{CpT} for $x\leq 1$), pointing
to the effect of the Ce-moment localization as the Kondo scale decreases.

The appearance of the NFL regime is closely related to the non-magnetic disorder introduced by the Sb
substitution and the resulting distribution in Kondo temperatures. Upon increasing $x$, the system
evolves toward more localized \textit{f}-moments as charge fluctuations decrease. The concomitant
increase in disorder results in a distribution of the hybridization strength between the
\textit{f}-moments and conduction electrons. LFL develops at temperatures below the lowest Kondo
scales of the distribution. The observed NFL behavior at $x = 1$ and 1.3 arises from a distribution
of Kondo temperatures that extends below the lowest temperatures
\cite{Dobrosavljevic92,Miranda96,Miranda97}. Such a scenario predicts a logarithmic divergence of
$C_\text{el}(T)/T$ and a $\rho(T) = \rho_0-aT$ dependence of the electrical resistivity, which we
both find below 1\,K in CePt$_4$Ge$_{11}$Sb [see Fig.\,\ref{CpT} for $C(T)/T$ and Fig.\,\ref{rho}e
for $\rho(T)$]. Within this scenario, applying a magnetic field removes the incoherent spin-flip
scattering below some energy scale related to the magnetic field. This is reflected in the negative
magnetoresistance near $x\approx 1$ (see Fig.\,\ref{rho}e). Furthermore, the Kondo-disorder scenario
gives a natural explanation for the extended Sb concentration range where the NFL behavior is
observed.

As $x$ increases further (beyond $x>1.3$), the Kondo-temperature distribution continues to decrease
in temperature.  This lowering enhances the tendency toward magnetism.  For $x=1.5$, no long-range
order develops but the maximum in resistivity indicates that scattering is suppressed below
$T_\text{max}$. Since this maximum cannot be related to Kondo coherence, it is most likely caused by
short-range order among the Ce moments. The resulting effective field induced by the short-range
order suppresses incoherent spin-flip scattering below some scale, reminiscent of the effect of an
external magnetic field (see Fig.\,\ref{rho}e), thereby leading to a LFL groundstate. This is indeed
observed. Finally, for $x\geq 1.9$, magnetic long-range order develops. Although the low temperature
$\gamma$ value of CePt$_4$Ge$_{10.1}$Sb$_{1.9}$ in the magnetically ordered state is still enhanced,
it is much smaller than that at $x = 1.5$ as expected for weakened Kondo correlations. For even
higher concentrations the estimated $\gamma$ seems to decrease further. This substantiates that the
Ce moments become even more localized with increasing substitution level and that the magnetism is of
local rather than itinerant nature.

In summary, in CePt$_4$Ge$_{12-x}$Sb$_x$ the evolution of a local moment AF state starting from an
intermediate-valent state is controlled by charge doping. While on increasing Sb concentration in
CePt$_4$Ge$_{12-x}$Sb$_x$ local Ce moments begin to develop, they are at the same time becoming
screened by the Kondo effect. The delicate relation of correlation effects together with a
distribution in hybridization strength results in an unconventional phase diagram. The observation of
a NFL regime in specific heat and electrical resistivity and the overall features of the phase
diagram can be well explained by an interplay of correlation effects and Kondo disorder.

We thank A.~Hewson and  S. Wirth  for useful discussions. S.~K. and F.~S. acknowledge support by the
National Science Foundation under Grant No.\ PHYS-1066293 and the hospitality of the Aspen Center for
Physics.


\begin{thebibliography}{30}
\bibitem[{rem()}]{remarkAs}
\bibinfo{note}{The substitution of Ge by As is difficult and results in
  multi-phase samples.}
\bibitem[{CEF()}]{CEFnote}
\bibinfo{note}{Note, that the $\Gamma_7$ doublet groundstate is in line with expectations for a $J= 5/2$
 state in a cubic enviroment.}
\end{thebibliography}
\end{document}